\documentstyle[11pt]{article}
\def\cE{\cal E}

\def\cL{{\cal L}}

\newfont{\goth}{eufm10 scaled \magstep1}

\def\a{\alpha}

\def\b{\beta}

\def\c{\gamma}\def\C{\Gamma}
\def\d{\delta}
\def\e{\epsilon}

\def\h{\eta}
\def\k{\kappa}
\def\L{\Lambda}

\def\s{\sigma}\def\S{\Sigma}

\def\th{\theta}

\def\beq{\begin{equation}}\def\eeq{\end{equation}}
\def\beqa{\begin{eqnarray}}\def\eeqa{\end{eqnarray}}
\def\barr{\begin{array}}\def\earr{\end{array}}
\def\x{\xi}

\def\del{\partial}
\def\ua{\underline{\alpha}}
\def\ub{\underline{\phantom{\alpha}}\!\!\!\beta}
\def\uc{\underline{\phantom{\alpha}}\!\!\!\gamma}

\def\ud{\underline\delta}

\def\una{\underline a}\def\unA{\underline A}
\def\unb{\underline b}\def\unB{\underline B}
\def\unc{\underline c}\def\unC{\underline C}
\def\und{\underline d}
\def\une{\underline e}

\def\unm{\underline m}\def\unM{\underline M}

\def\nab{\nabla}

\let\la=\label

\let\bm=\bibitem

\def\uM{{\underline M}}
\def\umu{{\underline \mu}}
\def\nn{\nonumber}
\def\bd{\begin{document}}
\def\ed{\end{document}}
\def\ba{\begin{array}}
\def\ea{\end{array}}
\def\bea{\begin{eqnarray}}
\def\eea{\end{eqnarray}}
\def\ft#1#2{{\textstyle{{\scriptstyle #1}\over {\scriptstyle #2}}}}
\def\fft#1#2{{#1 \over #2}}
\newcommand{\be}{\begin{equation}}
\newcommand{\ee}{\end{equation}}
\newcommand{\eq}[1]{(\ref{#1})}
\def\eqs#1#2{(\ref{#1}-\ref{#2})}
\def\det{{\rm det\,}}
\def\tr{{\rm tr}}
\newcommand{\ho}[1]{$\, ^{#1}$}
\newcommand{\hoch}[1]{$\, ^{#1}$}
\def\ra{\rightarrow}
\def\uha{{\hat {\underline{\a}} }}
\def\uhc{{\hat {\underline{\c}} }}

\newcommand{\tamphys}{\it\small Center for Theoretical Physics,
Texas A\&M University, College Station, TX 77843, USA}

\newcommand{\sissa}{\it\small International School for Advanced Studies
 (SISSA/ISAS), Via Beirut 2, 34014 Trieste, Italy}

\newcommand{\kings}{\it\small Department of Mathematics, King's College,
London, UK}

\newcommand{\newton}{\it\small Isaac Newton Institute for Mathematical Sciences,
Cambridge, UK}

\newcommand{\auth}{\large C.S. Chu\hoch{1} and
E. Sezgin\hoch{2\dagger} }

\begin{document}

\hfill{SISSA 128/97/FM}

\hfill{CTP TAMU-39/97}

\hfill{hep-th/9710223}

\vspace{20pt}

\begin{center}

{\Large\bf M-Fivebrane from the Open Supermembrane}
\vspace{30pt}

\auth

\vspace{15pt}

\begin{itemize}
\item[$^1$] {\small \em
International School for Advanced Studies (SISSA),
Via Beirut 2, 34014 Trieste, Italy}
\item[$^2$] {\small \em Center for
Theoretical Physics, Texas A\&M University, College Station, TX
77843, USA}
\end{itemize}

\vspace{60pt}

{\bf Abstract}

\end{center}

Covariant field equations of M-fivebrane in eleven dimensional curved
superspace are obtained from the requirement of $\k$-symmetry of an
open supermembrane ending on a fivebrane. The worldvolume of the
latter is a $(6|16)$ dimensional supermanifold embedded in the $(11|32)$
dimensional target superspace. The $\kappa$-symmetry of the system
imposes a constraint on this embedding, and a constraint on a modified
super 3-form field strength on the fivebrane worldvolume. These
constraints govern the dynamics of the M-fivebrane.

{\vfill\leftline{} \vfill
\vskip	10pt
\footnoterule
{\footnotesize \hoch{\dagger} Research supported in part by NSF Grant
PHY-9722090 \vskip -12pt}
\vskip	10pt

\pagebreak
\setcounter{page}{1}

\section{Introduction}

Just as open strings can end on D-branes, an eleven dimensional open
supermembrane can end on M-fivebrane. This possibility was first
considered in \cite{as,pkt1}. Further aspects of the eleven
dimensional open supermembrane were studied recently in
\cite{BB,mc,brax1,ezawa,brax2}. In particular, it has been suggested in
\cite{ezawa} that the $\k$-symmetry of the open supermembrane ending on
a M-fivebrane may give rise to the M-fivebrane equations of motion. In
this paper we will show that this is indeed the case.

In the model we consider, the worldvolume of the M-fivebrane is taken to
be a supersubmanifold, $M$, of the eleven dimensional target superspace,
$\uM$. The supermembrane action is an integral over a bosonic three
dimensional worldvolume $\S$, with its boundary $\del\S$ embedded in the
supermanifold $M$, such that
\be
\del \S \subset M \subset {\unM} \la{ms}
\ee

The requirement of $\kappa$-symmetry of the open supermembrane is shown
to have the following consequences. Firstly, $\kappa$-symmetry on $\S$
requires that the eleven dimensional supergravity equations are
satisfied. Moreover, the $\k$-symmetry on the boundary, which is the odd
diffeomorphisms of $M$ restricted to $\del\S$, imposes a constraint on
the embedding of $M$ in ${\unM}$ which says that the odd tangent space
of the worldsurface at any point is a subspace of the odd tangent space
of the target space at the same point. Furthermore, $\kappa$-symmetry on
$\del\S$ imposes a constraint on a modified super three-form field
strength $H$ defined as
\be
H=dB-f^*C\ ,
\ee
where $B$ is the super 2-form potential on the superfivebrane
worldvolume, $M$, and $f^*C$ is the pullback of the target space super
3-from $C$ to $M$. The superembedding constraint and the $H$-constraint
determine completely superfivebrane equations of motion.

In the superembedding approach to the description of superbranes as
emphasized in \cite{hs1,hs2}, the superembedding equation played a
central role. In the case of M-fivebrane, the
3-form $H$ was introduced for convenience in describing the field
equations and it was shown that the $H$-constraint is a consequence of
the superembedding condition \cite{hsw1,hsw2}. In the approach presented
in this paper, both the superembedding condition and the $H$-constraint
arise naturally from the requirement of $\k$-symmetry. The virtue of our
approach becomes more apparent when we apply the formalism to super
$D$-branes. In that case, one finds that the superembedding condition is
not sufficient by itself to imply the super $D$-brane equations of
motion, but one needs the analog of the $H$-constraint to do so, at
least for $p\ge 6$ \cite{hssw}. Thus, it is remarkable that the
considerations of the $\k$-symmetry of the open branes ending on other
branes naturally give all the constraints needed to describe the
dynamics of the total system.

The nature of the supersubmanifold $M$ may appear to be put in by hand.
However, the $\kappa$-symmetry is powerful enough to restrict
the nature of the possible supersubmanifolds. In particular, it is
known that the $(6|16)$ dimensional submanifold is allowed
\cite{hs1,hs2}. We use the notation $(D|D')$, where $D$ is the real
bosonic dimension and $D'$ is the real fermionic dimension of a
supermanifold. The possibility of a $(10|16)$ dimensional submanifold
has been conjectured in \cite{hs1}, and the possibility of a $(2|16)$
dimensional submanifold has been pointed out in \cite{ezawa}. The
determination of whether such configurations exist requires further
analysis.

\section{Open Supermembrane Ending on Superfivebrane}

The eleven dimensional supermembrane was studied in \cite{bst1,bst2}. In
this section, we will study an open supermembrane $\S$ with its boundary
$\del \S$ couple to a 2-form superfield. For simplicity, we will take
$\del\S$ to have a single boundary component. The membrane worldvolume
is bosonic. We will take its boundary, however, to lie in a bosonic
submanifold of a supermanifold $M$ of dimension $(6|16)$, which in turn
is a submanifold of a target space $\unM$ of dimension $(11|32)$. We use
the notations and conventions of \cite{hs1}. In particular, we denote by
$z^{\unM}=(x^{\unm},\th^{\umu})$ the local coordinates on $\unM$, and
$A=(a,\a)$ is the target tangent space index. We use the ununderlined
version of these indices to label the corresponding quantities on the
worldsurface. The embedded submanifold $M$, with local coordinates
$y^M$, is given as $z^{\unM}(y)$.

We consider the following action for an open supermembrane ending on a
superfivebrane
\be
S=-\int_\S d^3 \xi \left ( \sqrt{-g} + \e^{ijk} C_{ijk}\right)
+ \int_{\del\S} d^2 \s \e^{rs} B_{rs}\ ,\la{action}
\ee
where $\x^i~(i=0,1,2)$ are the coordinates on the membrane worldvolume
$\S$, $\s^r~(r=1,2)$ are the coordinates on the boundary $\del\S$,
$g_{ij}$ is the metric on $\S$ and $g=\det g_{ij}$.

In addition to the usual super 3-form $C$ in $(11|32)$ dimensional
target superspace $\uM$, we have introduced a super 2-form $B$ on the
$(6|16)$ dimensional superfivebrane worldvolume $M$, which is a {\it
supersubmanifold} of $\uM$. The suitable pullbacks of these superforms,
and the induced metric occuring in the action are defined as:
\bea
C_{ijk} &=& E_i{}^{\unA} E_j{}^{\unB} E_k{}^{\unC} C_{\unC\unB\unA}\ ,
\nn\\
B_{rs} &=& E_r{}^A E_s{}^B B_{BA}\ ,
\nn\\
g_{ij} &=& E_i{}^{\una} E_j{}^{\unb} \eta_{\una\unb}\ ,\la{cbg}
\eea
where $\eta_{\una\unb}$ is the Minkowski metric in eleven dimensions, and
\bea
E_i{}^{\unA} &=& \del_i z^{\uM} E_{\uM}{}^{\unA}\ ,
\nn\\
E_r{}^A &=& \del_r y^M E_M{}^A \ ,
\la{ee}
\eea
where $E_{\uM}{}^{\unA}$ is the target space supervielbein and
$E_M{}^A$ is the worldsurface supervielbein.
Defining the basis one-forms $E^{\unA} = d\xi^i E_i{}^{\unA}$ and
$E^A = d\s^r E_r{}^A$, note the useful relation
\be
E^{\unA}|_{\del\S} = E^A E_A{}^{\unA}|_{\del\S}\ .\la{useful}
\ee

The embedding matrix $E_A{}^{\unA}$ plays an important role in the
description of the model, and it is defined as
\be
E_A{}^{\unA}=E_A{}^M\del_{M}z^{\unM}E_{\unM}{}^{\unA},
\ee

The action \eq{action} is invariant under diffeomorphisms of $\S$, with
suitable boundary conditions imposed on the parameter of the
transformation, as well as the tensor gauge transformations
\bea
&& \d C= d \L\ , \nn\\
&& \d B= f^*\L\ ,  \la{gt}
\eea
where $\L(z^{\unM})$ is a super 2-form in $\uM$, and the pullback $f^*\L$ of a
vector $V$ on $\unM$ to $M$ is defined as
\be
(f^* V)_A = E_A{}^{\unA}
V_{\unA}\ .
\ee

We shall now require the total action to be invariant under the
$\k$-symmetry transformation. On  the interior of $\S$,
they take the usual form \cite{bst1}
\bea
\d_{\k} z^{\una} &=& 0 \ , \la{k1}\\
\d_{\k} z^{\ua} &=&  \k^{\uc}(\xi) (1+\C_{(2)})_{\uc}{}^{\ua} \ ,\la{k2}
\eea
where
\be
\d_\k z^{\unA} = \d_\k z^{\uM} E_{\uM}{}^{\unA}\ , \la{v1}
\ee
and
\be
\C_{(2)} = \frac1{3!\sqrt{-g}} \e^{ijk} \c_{ijk}\ ,\la{C}
\ee
where the pullback $\c$-matrices are defined as
\be
\c_i = E_i{}^{\una} \C_{\una}\ .
\ee

We also need to specify the fermionic $\k$-symmetry
transformations of $z^{\unA}$ on the boundary $\del\S$.
Without loss of generality, they take the form
\bea
\d_{\k} z^{\una} &=& 0\ , \la{bgk1}\\
\d_{\k} z^{\ua} &=& \k^{\uc}(\s) P_{\uc}{}^{\ua} \quad\quad
\mbox{on $\del\S$} \ , \la{bgk2}
\eea
where $P_{\uc}{}^{\ua}$ is some projector, whose explicit form will be
spelled out later (see \eq{c5}).

We next derive an interesting consequence of the $\k$-transformations
specified above. To do so, we first observe that an arbitrary
transformation of $y^M$ induces a transformation on $z^{\unM}$ given by
\be
\d z^{\unA} = \d y^A E_A{}^{\unA}\ \quad\quad \mbox{on $M$},
\la{dzb}
\ee
where
\be
\d y^{A} = \d y^{M} E_{M}{}^{A}\ . \la{v2}
\ee

It follows from \eq{bgk1} and \eq{dzb}
that $\d_\k y^a$ and $\d_\k y^{\a}$ satisfy
\be
0= \d_\k y^a E_a{}^{\una}+  \d_\k y^{\a} E_{\a}{}^{\una}\ , 
\la{dz2}
\ee
on the boundary $\del \S$. The $\una=b$ component of this equation is $0=
\d_\k y^a E_a{}^b+ \d_\k y^{\a} E_{\a}{}^b$. One can check that
$E_\a{}^b$ can be gauged away by using the bosonic diffeomorphisms of
$M$, namely $\d_\eta y^M E_M{}^a=\eta^a$. Hence, one can set $E_\a{}^b=0$,
and since $E_a{}^b$ is invertable, it follows that
\be
\d_\k y^a=0\ , \la{dza0}
\ee
on $\del\S$, and hence on $M$. Using this in $\una=b'$ component of
\eq{dz2}, and observing that $\d_\k y^\a$ is an arbitrary odd
diffeomorphism of $M$, it follows that $E_\a{}^{b'}=0$. Recalling that
$E_\a{}^b=0$ as well, we get
\be
E_{\a}{}^{\una}=0\ .  \label{basic}
\ee
This is the superembedding condition that plays a crucial role in the
description of superbrane dynamics \cite{hs1,hs2,hsw1}.

We found the $\k$-transformation $\d_\k y^a$ on $M$ above. We now turn
to the determination of the remaining variation $\d_\k y^\a$ on $M$.
Using \eq{dza0} and \eq{basic} in \eq{dzb}, we find
\be
\d_\k y^\a E_\a{}^{\ua}= \d_\k z^{\ua}\ ,\la{bk2}
\ee
on the boundary ${\del\S}$. To solve for $\d_\k y^\a $, it is useful to
introduce a normal basis $E_{A'}=E_{A'}{}^{\unA} E_{\unA}$ of vectors at
each point on the worldsurface. The inverse of the pair
$(E_A{}^{\unA},E_{A'}{}^{\unA})$ is denoted by
$(E_{\unA}{}^A,E_{\unA}{}^{A'})$ \cite{hsw1}. From \eq{bk2}, it follows
that
\be
\d_{\k} y^{\a}= \d_\k z^{\ua} E_{\ua}{}^{\a}\ ,
\ee
on the boundary ${\del\S}$. This means that the variation $\d_\k y^\a$
is an arbitrary odd-diffeomorphism, effecting the 16 fermionic
coordinates of $M$, and that when restricted to $\del\S$, it agrees with
the $\kappa$-symmetry transformation on ${\unM}$, which also has 16
independent fermionic parameters.

Now we are ready to seek the conditions for the $\k$-symmetry of the
action \eq{action}
\footnote
{
The $\k$-symmetry of the action \eq{action} in a flat target superspace
was also considered in \cite{brax1}, where the consequences of the
resulting constraints are not considered. Moreover, our results for the
constraints differ from theirs.
}.
Using \eq{dzb} and \eq{dza0} in the variation of the
action, we find that the vanishing of the terms on $\S$ imposes
constraints on the torsion super 2-form $T$ and the super 4-form
$G=dC$, such that they imply the equations of motion of the eleven
dimensional supergravity \cite{bst1}. The non-vanishing parts of the
target space torsion are \cite{cf,bh}
\bea
T_{\ua\ub}{}^{\unc} &=& -i(\C^{\unc})_{\ua\ub}\ ,
\nn\\
T_{\una\ub}{}^{\uc} &=&-
{1\over36}(\C^{\unb\unc\und})_{\ub}{}^{\uc}G_{\una\unb\unc\und}
-{1\over288}(\C_{\una\unb\unc\und\une})_{\ub}{}^{\uc}
G^{\unb\unc\und\une}\ , \la{tt}
\eea
and $T_{\una\unb}{}^{\uc}$. The only other non-vanishing component of
$G$ are
\be
G_{\una\unb\uc\ud}=-i(\C_{\una\unb})_{\uc\ud}\ .
\ee

The remaining variations are on the boundary, and yield the final result
\be
\d_\k S=\int_{\del \S} \e^{rs} E_r^{A} E_s^{B} \d_\k y^\c  H_{\c B A}\ ,
\la{var}
\ee
where
\be
H=dB-f^{*} C\ ,
\ee
and satisfies the Bianchi identity\footnote{
The two-form $B$ has to be rescaled by a factor of four to agree with
the conventions of \cite{hsw1}.}
\be
dH=-f^{*} G\ .  \la{bi1}
\ee
Since $\d_\k y^\a$ are arbitrary, the vanishing of \eq{var} implies the
constraint
\be
H_{\c B A}=0\ .\la{h}
\ee
Thus the only nonvanishing component of $H$ is $H_{abc}$. The
constraints \eq{basic} and \eq{h} encode elegantly all the information
on the superfivebrane dynamics, as has been shown in
\cite{hs1,hs2,hsw1}. For completeness, we have collected in the Appendix
the covariant superfivebrane equations of motion which follow from these
constraints
\footnote{See \cite{pst,sch} for the $M$-fivebrane action in the
Green-Schwarz formalism, and \cite{bandos} for its relation to 
\cite{hs1,hs2,hsw1}.
}.
It should be noted that the $H_{abc}$ component does not participate in
\eq{h}. Thus, one has the freedom to envisage a cusp-like behaviour in
$B_{ab}$ giving rise to a discontinuity in $(dB)_{abc}$. Taking into
account this discontinuity, the $abcd$ component of the Bianchi identity
\eq{bi1} gets modified as
\footnote
{
In \cite{brax2}, a modified Bianchi identity of this kind is derived by
adding a bosonic piece of the M-fivebrane action to \eq{action}.
However, the $\k$-symmetry of this system is by no means clear.
}
\be
dH=-f^{*} G +  \d_W\ ,  \la{bi2}
\ee
where $ \d_W$ is the Poincar\'e dual of $W=\del\S$ in the bosonic
fivebrane worlvolume $\S_6$. A similar argument applies to the
background field $C_{\una\unb\unc}$, thereby modifying the
$\una\unb\unc\und\une$ component of the Bianchi identity
\be
dG=\d_W\ , \la{w}
\ee
where now $ \d_W$ is the Poincar\'e dual of $W=\S_6$ in the eleven
dimensional bosonic target space $Q$ \cite{witten}.

The modification \eq{bi2} is important for the analysis of the
reparametrization anomalies localized on $\del\S$, and \eq{w} is
relevant for the reparametrization anomalies on the fivebrane
\cite{da,chu}. However, these modifications are not essential for the
purposes of this paper, where we derive the superfivebrane equations of
motion in the bulk of its worldvolume. Hence, we shall drop the $\d_W$
terms in the rest of this paper.

It is known that the superembedding condition \eq{basic} implies the
$H$-constraint \eq{h}. It would be interesting to determine if the
reverse is true. To this end, we have examined the the $\a\b\c d$
component of the Bianchi identity in flat target superspace and at the
linearized level in the fivebrane worldvolume fields. Interestingly, we
find that the Bianchi identity \eq{bi1} indeed implies the embedding
condition at this level.

To conclude this section, we discuss the nature of the global supersymmetry
of the model in a flat target superspace. Consider the
transformations
\be
\d_\e z^{\ua} = \e^{\ua}\ ,\la{susy}
\ee
where $\e^{\ua}$ is a constant parameter.
Due to \eq{dzb}, one finds that \eq{susy} induces
a transformation on $M$ satisfying
\be
\d_\e y^\a E_\a{}^{\ua}= \e^{\ua}\ .\la{bk3}
\ee
It should be emphasized that these
transformations, as well as the $\k$-symmetry transformations are not
special cases of the $\L$-transformations. The latter is a
transformation of the background fields involving a parameter that is an
arbitrary function of the target space coordinates.

Substituting the variation \eq{bk3}, we find that the action \eq{action}
is invariant, upon the use of the $H$-constraint \eq{h}. It remains to
analyse the consequences of the condition \eq{bk3}. Multiplying \eq{bk3}
with $E_{\ua}{}^{\b'}$ (defined earlier), we get
\be
\e^{\ua}E_{\ua}{}^{\b'}= 0\ .\la{bk4}
\ee
The target space spinor index $\ua$, running
fom 1 to 32 is split in two, $\ua\rightarrow(\a,\a')$, where both the
worldsurface index $\a$ and the normal index $\a'$ run from 1 to 16.

To elucidate the meaning of this condition on the parameter $\e^{\ua}$,
it is useful to define the projection operators
\bea
E_{\ua}{}^{\a} E_{\a}{}^{\uc} &=& \ft12 (1+\C_{(5)})_{\ua}{}^{\uc}\ , \nn\\
E_{\ua}{}^{\a'} E_{\a'}{}^{\uc} &=& \ft12 (1-\C_{(5)})_{\ua}{}^{\uc}\ , \la{p2}
\eea
where $\C_{(5)}$ satisfies $\C_{(5)}^2=1$. This matrix is defined by
\eq{p2} and is given in the Appendix. Multiplying \eq{bk4} with
$E_{\b'}{}^{\ub}$ gives
\be
\e^{\ua}(1-\C_{(5)})_{\ua}{}^{\ub}= 0\ . \la{sc}
\ee
This means that at most half of the
supersymmetry can survive. Furthermore, since the matrix $\C_{(5)}$
defined in \eq{af} is a complicated function of the worldvolume fields,
the condition \eq{sc} severely restricts the allowed configurations for
them. One possibility is to set all the worldvolume fields equal to
zero. In that case, and in a physical gauge, $\C_{(5)}$ becomes a
product of constant worldvolume $\c$-matrices, implying that half of
target space supersymmetry.

Finally, we note that just as $\e^{\ua}$ satisfies \eq{sc}, the
variation $\d_\k z^{\ua}$ given in \eq{bgk2} satisfies $\bar{\k} P
(1-\C_{(5)}) = 0$ on the boundary $\del \S$. This can be seen by
multiplying the $\unA=\ua $ component of \eq{dzb} by $E_{\ua}{}^{\a'}
E_{\a'}{}^{\uc}$ and noting that $E_{\a}{}^{\ua} E_{\ua}{}^{\b'}=0$.
This condition can be satisfied by taking
\be
P = \ft12 (1+\C_{(5)}).
\la{c5}
\ee

\section{Boundary Conditions}

In this secton we consider the boundary conditions that arise from the
variation of the action \eq{action}. The requirement of the action be
stationary when the supermembrane field equations of \cite{bst2} hold
imposes the boundary condition
\be
\int_{\del\S} ( \d y^A E_A{}^{\una} \sqrt{-g} n^i E_{i\una}
+ \d y^C n_i\e^{ijk} E_j{}^B E_k{}^A H_{ABC})=0 \ . \la{bt1}
\ee
Using \eq{basic} and \eq{h}, we obtain
\be
\int_{\del\S} [ \d z^{a'}\sqrt{-g} n^i E_{i a'} +
\d y^c (\sqrt{-g} E_c{}^{\una} n^i E_{i\una} +n_i\e^{ijk} E_j{}^b E_k{}^a
H_{abc} ) ]=0 \ , \la{bt2}
\ee
where $\d z^{a'}$ is defined in \eq{v1}. This equation is satisfied by
imposing the Dirichlet boundary condition,
\be
\d z^{a'}|_{\del\S} =0\ ,
\ee
and the Neumann boundary condition
\be
\left( \sqrt{-g} n^iE_{ic}+n_i\e^{ijk} E_j{}^a E_k{}^b H_{abc}
\right)|_{\del\S}=0\ , \la{dbc}
\ee
where $n^i$ is a unit vector normal to the boundary $\del\S$, and $a'$
labels the directions transverse to the fivebrane worldvolume. We expect
that this constraint is $\k$-invariant, modulo the fermionic field
equation of the superfivebrane (see the Appendix) \cite{hsw1}
\be
{\cE}_a(1-\C_{(5)})\c^b m_b{}^a=0\ . \la{Dirac}
\ee

A boundary condition similar to \eq{dbc} has also been discussed in
\cite{ezawa} for a flat target space and a purely bosonic 2-form on the
fivebrane worldvolume. These authors set the $C$-dependent term of $H$
in \eq{dbc} equal to zero separately by imposing suitable boundary
condition on the fermionic variables, involving a projector of the form
$(1+\C_{(5)})$ with $\C_{(5)}$ defined as in \eq{af}.

In \cite{BB}, on the other hand, the 2-form field $B$ is not considered,
and the $C$-dependent terms are set equal to zero at the boundary, by
projecting the fermionic variables in particular gauges. In this case,
one has to check that the boundary terms due to the global supersymmetry
variation of the supermembrane action vanish, and indeed they do
\cite{BB}. It should be emphasized that this is possible only by
sacrificing the eleven dimensional super Poincar\'e invariance
\cite{bst2}. For example, the boundary terms cannot be made to vanish
for an open supermembrane with boundaries moving freely in ${\unM}$.

For completeness, we also consider the boundary terms due to
the reparametrization transformations
\beq
\d z^{\unM} = v^i \del_i z^{\unM},
\eeq
under which the action transforms as
\be
\d S =\int_{\S} d^3 \x \del_i(v^i \cL)\ .
\eeq
This boundary term vanishes by imposing the condition
\be
n_i v^i |_{\del\S} =0 \ .
\ee
Requiring that the boundary condition \eq{dbc} is preserved by
the reparametrization transformations imposes the conditions
\be
n^i \del_i v^r|_{\del\S} =0\ .
\ee

\section{Comments}

In this paper we considered an open supermembrane with a bosonic
worldvolume $\S$ ending on a superfivebrane with worldvolume $M$, which
is a $(6|16)$ dimensional supersubmanifold of the $(11|32)$ dimensional
target superspace ${\unM}$. We showed that the requirement of $\k$-
symmetry not only constraints the eleven dimensional curved background
fields to satisfy their equations of motion, but also yields a
superembedding condition and a constraint on a modified 3-form field
strength, $H$,  which determine the superfivebrane equations of motion.

Our formalism can be applied to other possible choices of embeddings
\eq{ms}. In particular there are two special cases which deserve further
analysis within the present framework. In one case, the supermembrane
ends on a superstring, and thus $\del \S = M_B$, where $M_B$ is the
bosonic part of the $(2|16)$ dimensional string superworldsheet. In
another case, the supermembrane ends on the boundary of ${\unM}$, namely
$M=\del {\unM}$, corresponding to a M-ninebrane. The latter case is
similar to the configuration considered in \cite{hw}. For a $\k$-symmetric
formulation of an open supermembrane ending on the Horawa-Witten ninebrane,
see \cite{mc,brax1}.

In a separate paper \cite{cs2}, we will show that the ideas presented
here apply also to fundamental type II strings ending on
$D$-branes, $D2$-brane ending on solitonic type IIA fivebrane, and
open $Dp_1$-branes ending on $Dp_2$-branes.

The application of our formalism to type I open branes, i.e. branes
in a target space with sixteen real supersymmetries, should also be
possible. It would be interesting, for example, to derive the equations
of motion for the heterotic fivebrane, via the study of heterotic
string ending on solitonic fivebrane.

Another possible generalization of the present work is to consider open
branes ending on branes that do not possess maximal transverse space
rotational symmetry. For example, the $pp$ waves and the sixbranes, also
known as Kaluza-Klein monopoles, in eleven dimensions can be considered
as possible end point surfaces for the eleven dimensional supermembrane.

One may also consider elevating the open brane worldvolumes considered
here to supermanifolds
\footnote{
We thank Paul Howe for a discussion about this possibility.
}.
Although an action is not known for such systems, one may nonetheless
obtain the equations of motion for all the branes involved by
generalizing the usual superembedding approach to deal with triplet
\eq{ms} of supermanifolds $\S$, $M$ and ${\unM}$.

Finally, a matrix regularization \cite{deWit1,deWit2,rey,ezawa} of
the action \eq{action} may be be relevant for the generalization of
M(atrix) theory \cite{bssf} in a curved background \cite{d1} and in the
presence of fivebrane \cite{d2}.

\vspace{0.75cm}

{\Large \bf Acknowledgements}

We are very grateful to Paul Howe for helpful e-mail communications. E.S.
would like to thank the International Center for Theoretical Physics for
hospitality.

\bigskip\bigskip

\noindent{\Large\bf Appendix}

\bigskip

\noindent{\large\bf $M$-Fivebrane Equations of Motion }

\bigskip

Here, we give the nonlinear field equations of the
superfivebrane equations, up to second order fermionic terms, that
follow from the superembedding condition \eq{basic}, which are proposed to
arise equally well from the $H$-constraint \eq{h}. These equations are
\cite{hs1,hs2,hsw1}:
\bea
&&
{\cE}_a(1-\C_{(5)})\c^b m_b{}^a=0\ , \la{Dirac4}\\
&&\nn\\
&&
G^{mn}\nab_m H_{npq} = {3\sqrt{-g} \over 128(1-\ft23 \tr\,k^2)}
\big(1-\ft23 k \big)^4{}_{[p}{}^n~\e_{q]nm_1\cdots m_4}~G^{m_1\cdots m_4}
\ ,\nn\\
&&
G^{mn}\nab_m{\cE}_n{}^{\unc} = {(1-\ft23 \tr\,k^2)\over 6!\sqrt{-g}}
\epsilon^{m_1\cdots m_6 }(
G^{\underline a}{}_{m_1\cdots m_6}
+ \ft23 G^{\una}{}_{m_1m_2m_3}\,H_{m_4m_5m_6}\,)
(\d_{\una}{}^{\unc}-{\cE}_{\una}{}^m{\cE}_m{}^{\unc})\ , \nn
\eea
where
\beqa
{\cE}_m{}^{\una}(x) &=& \del_m z^{\unM} E_{\unM}{}^{\una}\qquad {\rm at}\
\th=0\ , \nn\\
{\cE}_m{}^{\ua}(x) &=& \del_m z^{\unM} E_{\unM}{}^{\ua}\qquad {\rm at}\ \th=0\ .
\eeqa
are the embedding matrices in the Green-Schwarz formalism. The matrix
$\C_{(5)}$ at $\theta =0$ is given by
\be
\C_{(5)}=- \left[{\rm exp}~ (-\ft13\c^{mnp}h_{mnp})\right] \C_{(0)}\ , \la{af}
\ee
where
\beq
\C_{(0)}={1\over 6!\sqrt {-g}} \e^{m_1\cdots m_6} \c_{m_1\cdots m_6}\ .
\la{c0}
\eeq
The pullback $\c$-matrices in \eq{af} and \eq{Dirac4} are defined by 
\be
\c_m=\C_{\una} {\cE}_m{}^{\una}\ , \quad\quad 
\c^b=\c^m e_m{}^b.
\ee
The  matrix $e_m{}^a$ is the vielbein for the induced metric 
\bea
g_{mn}(x) &=& {\cE}_m{}^{\una}{\cE}_n{}^{\unb}\h_{\una\unb}\nn\\
&=& e_m{}^a e_n{}^b \eta_{ab}\ ,
\eea
and $G^{mn}$ is another metric defined as
\be
G^{mn}  =(m^2)^{ab}e_a{}^m e_b{}^n\ ,
\ee
where 
\be
m_a{}^b=\d_a{}^b-2k_a{}^b,
\ee
\be
 k_a{}^b = h_{acd} h^{bcd}\ ,
\ee
and $h_{abc}$ is 
a self-dual field strength which is related to $H_{abc}$ via the
equation
\be
h_{abc}=m_a{}^d H_{cde}\ .
\ee
The $G_7$ is the seven form that occurs in the dual formulation of
eleven dimensional supergravity,
\beq
G_{\underline d_1\ldots \underline d_4}= {1\over 7!}
\epsilon _{\underline d_1\ldots \underline d_4
\underline e_1\ldots \underline e_7}
G^{\underline e_1\ldots \underline e_7}.\
\eeq
The
target space indices on $G_4$ and $G_7$ have been converted
to worldvolume indices with factors of ${\cE}_m{}^{\una}$.

The $\k$-symmetry transformation rules are
\bea
\d_{\k} z^{\una} &=& 0 \ , \nn\\
\d_{\k} z^{\ua} &=&  \k^{\uc} (1+\C_{(5)})_{\uc}{}^{\ua} \ ,\nn\\
\d_\k h_{abc} &=& -\ft{i}{16} m_{[a|}{}^d\,{\cE}_d(1-\C_{(5)})\c_{|bc]} \k\ ,
\la{kt}
\eea
where $\C_{(5)}$ is given by \eq{af}.

\pagebreak

\ed